# Symmetry Origins of the Field-Free Superconducting Diode Effect in the Kagome Superconductor $CsV_3Sb_5$

Xin-Jie Liu[1,2,10], Shengbiao Sun[2,3,10], Ke-Fan Song[2,3,10], Jia-Peng Peng[2,3,10], Xilin Feng[4], Lang Xiao[2,3], Tong Liu[5], Qilin Han[6], Ya-Qing Bie[6], Ning Kang[7], Xiaosong Wu[8], Yanfei Wu[1*], Shouguo Wang[1,9*], Kam Tuen Law[4], Shuo Wang[2*], Dapeng Yu[2], and Ben-Chuan Lin[2*]

[1]*School of Materials Science and Engineering, Key Laboratory of Advanced Materials and Devices for Post-Moore Chips, Ministry of Education, University of Science and Technology Beijing, Beijing 100083, China.*

[2]*International Quantum Academy, and Shenzhen Branch, Hefei National Laboratory, Shenzhen 518048, China.*

[3]*Southern University of Science and Technology, Shenzhen 518055, China.*

[4]*Department of Physics, Hong Kong University of Science and Technology, Clear Water Bay, Hong Kong 999077, China.*

[5]*School of Physics, Ningxia University, Yinchuan 750021, China.*

[6]*State Key Laboratory of Optoelectronic Materials and Technologies, Sun Yat-sen University, Guangzhou 510275, China.*

[7]*Key Laboratory for the Physics and Chemistry of Nanodevices, School of Electronics, Peking University, Beijing 100871, China*

[8]*State Key Laboratory for Artificial Microstructure and Mesoscopic Physics, Frontiers Science Center for Nano-optoelectronics, Peking University, Beijing 100871, China.*

[9]*Anhui Provincial Key Laboratory of Magnetic Functional Materials and Devices, School of Materials Science and Engineering, Anhui University, Hefei 230601, China.*

[10]*These authors contributed equally to this work.*

[*]Corresponding author. Email: yanfeiwu@ustb.edu.cn; sgwang@ahu.edu.cn; wangshuo@iqasz.cn; linbenchuan@iqasz.cn

# Abstract

Field-free superconducting diode effects require both inversion-symmetry breaking and an internal time-reversal-symmetry (TRS) breaking field, making them sensitive probes of hidden order in superconductors. In centrosymmetric kagome $AV_3Sb_5$, the inversion symmetry generally should generally preclude the observation of the superconducting diode effect. Furthermore, though TRS breaking has been reported in the superconducting regime of $CsV_3Sb_5$, whether it is generated by superconductivity or inherited from charge-density-wave (CDW) order remains unresolved. Here we show that pristine $CsV_3Sb_5$ devices exhibit no intrinsic field-free superconducting diode effect, whereas surface oxidation or asymmetric etching activates a large nonreciprocal supercurrent. Moreover, the response is stochastic, with sweep-dependent polarity and magnitude, indicating metastable TRS-breaking domain configurations. Small out-of-plane magnetic fields stabilize the superconducting diode response, consistent with field selection of such domains. Finally, when long-range CDW order is suppressed by Ti doping, the SDE disappears. Our results establish the symmetry requirements for the field-free SDE in $CsV_3Sb_5$, reveal its stochastic domain-controlled character, and link superconducting-state TRS breaking to CDW-related order.

The superconducting diode effect (SDE)[1–3], namely the nonreciprocal flow of supercurrent without energy dissipation, has recently emerged as a central topic in superconducting electronics and quantum devices. Although most demonstrated SDEs require an applied magnetic field[4–9], the field-free counterpart is particularly compelling from both technological and fundamental perspectives. Eliminating the need for external magnetic fields simplifies device integration, reduces operational complexity, and is advantageous for scalable superconducting circuits. At the same time, a finite field-free SDE imposes stringent symmetry requirements, indicating the coexistence of inversion-symmetry breaking and time-reversal-symmetry breaking within, or coupled to, the superconducting state. Specifically, the SDE in a pristine superconductor is proportional to $\left(\hat{P} \times \hat{B}\right) \cdot \hat{I}_c$, where the polar axis $\hat{P}$, the effective magnetic field $\hat{B}$ associated with TRS breaking, and the current direction $\hat{I}_c$ are mutually orthogonal[3]. Thus, in principle, the SDE should only be observed in noncentrosymmetric superconductors that break time-reversal symmetry. Field-free SDE has been achieved in magnetic heterostructures or systems incorporating ferromagnetic atomic layers[10–13], where the magnetic components break time-reversal symmetry. Only a few reports describe field-free SDE without magnetic constituents, which may be attributed to finite-momentum Cooper pairing or unconventional order parameters that spontaneously break time-reversal symmetry[14–18]. Field-free superconducting nonreciprocity can thus serve not only as a device functionality but also as an electrical probe of hidden symmetry breaking[3].

$AV_3Sb_5$[19–23] have recently emerged as a fertile platform for exploring correlated quantum phenomena. These materials crystallize in the centrosymmetric space group *P6/mmm* and exhibit an unusual electronic structure featuring Dirac cones, van Hove singularities, and flat bands. Their normal state hosts charge-density-wave order[24–39] and electronic nematicity[30,31,35,36], while the superconducting state displays a remarkable diversity, including possible singlet or multiband behavior[40–43], pressure-induced double-dome phases[32,44], pair-density-wave correlations[45–48]. The TRS breaking has also been reported in the CDW phase of the normal states by magneto-optical Kerr effect [31] and μSR measurements[29,49,50]. It has also been inferred to exist in the superconducting state from superconducting diode or Josephson signatures[17], μSR[49] and scanning tunneling microscopy measurements[46,47,49]. These observations establish $AV_3Sb_5$ as an important candidate of unconventional superconductors[51], but they leave a central question unresolved: does the TRS breaking observed in the superconducting regime originate from a TRS-broken superconducting order parameter, or is it inherited from a CDW-related TRS-broken electronic texture that persists into, and couples to, the superconducting state?

The centrosymmetry of $AV_3Sb_5$ complicates a direct diode-based test of TRS breaking. Even though CDW and nematic orders reconstruct the electronic structure and reduce rotational symmetry, global inversion symmetry is expected to remain intact in pristine $AV_3Sb_5$ at ambient pressure[30–32,35,36]. A pristine device can therefore contain an internal TRS-breaking electronic order and still show no field-free superconducting diode effect, because the polar axis required for superconducting nonreciprocity is absent. Conversely, observing a field-free diode response in $AV_3Sb_5$ requires an additional source of inversion-symmetry breaking, which in thin devices can arise from surfaces or edges, oxidation, fabrication-induced asymmetry and other effects. A controlled experiment must therefore first establish the centrosymmetric baseline, then deliberately introduce inversion symmetry breaking, and finally ask whether an internal TRS-breaking order is present to couple to that extrinsic polar environment.

Here we report such a controlled transport study of $CsV_3Sb_5$ devices. hBN-encapsulated, air protected flakes provide the pristine reference and show no superconducting diode effect. Only by extrinsic inversion symmetry breaking, such as surface oxidation and asymmetric etching, then a field-free diode response could be observed. By conducting repeated current sweeps rather than relying on a single-shot measurement, we find that the activated SDE has stochastic polarity and magnitude, but can be trained by small out-of-plane magnetic fields. We then apply the same inversion-breaking protocols to Ti-doped $CsV_{3-x}Ti_xSb_5$ ($x$ = 0.24), where long-range CDW order is suppressed, and find no diode response. Together, these measurements establish an experimental sequence consisting of a pristine baseline, extrinsic inversion symmetry breaking, stochastic diode responses, field training, and a CDW-suppressed control, which points to TRS breaking nature of the superconducting states probably inherited from CDW orders.

# Results

## Pristine Device Characterization

The crystal structure of $AV_3Sb_5$ consists of alternating alkali, vanadium kagome, and antimony layers, with an inversion center at the Sb site in the kagome layer (Fig. 1a&b). To test whether an ideal thin flake has an intrinsic diode response, we fabricated $CsV_3Sb_5$ devices in an argon glovebox and encapsulated them with hBN. A representative device is shown in Fig. 1c. Its resistance drops into a zero-resistance state with $T_c^{onset}$ = 4.08 K and $T_c^{zero}$ = 3.39 K (Fig. 1d).

Differential resistance measurements at 1.8 K show overlapping positive- and negative-bias switching curves (Fig. 1e). For direct comparison, the negative-bias branch of the backward sweep is inverted and plotted against the forward sweep in Fig. 1F. The two traces coincide within experimental resolution, demonstrating that $|I_c^+|$ and $|I_c^-|$ are indistinguishable at zero magnetic field. The same conclusion holds across ten hBN-encapsulated $CsV_3Sb_5$ devices with different thicknesses and fabrication histories (Figs. S1 and S2 and Table S1). Thus, protected $CsV_3Sb_5$ obeys the centrosymmetric selection rule: without externally broken inversion symmetry, it does not exhibit an intrinsic field-free superconducting diode effect.

## Extrinsic inversion symmetry breaking and stochastic diode responses

We next introduced inversion-symmetry breaking in a controlled manner. In Device 11, the pristine hBN-encapsulated sample first showed no diode asymmetry (Fig. 2a). After removing the hBN cover and exposing the flake to ambient air, the differential-resistance peak evolved from a single transition into a multi-peak structure, and the critical current decreased from approximately 115 $\mu$A to approximately 37 $\mu$A (Fig. 2b). Such changes are consistent with surface oxidation and the superconducting inhomogeneity expected in disordered low-dimensional superconductors [52–54]. With further oxidation, a clear field-free diode response developed (Fig. 2c&d).

We quantify the diode response using

$$\eta = \frac{|I_c^+| - |I_c^-|}{|I_c^+| + |I_c^-|} \tag{1}$$

where $|I_c^+|$ and $|I_c^-|$ are defined as the onset currents of the superconducting-to-resistive transition in the forward and backward sweeps. In the second oxidation step of Device 11, $\eta$ reaches 83.61% (Fig. 2C). Continued oxidation suppresses the critical current and smears the transition, indicating that excessive disorder eventually degrades the superconducting channel.

A second, independent route to inversion-symmetry breaking is geometric asymmetry. We patterned Device 13 into an irregular pattern by selective etching (Fig. 3a, inset). A field-free superconducting diode effect is again observed (Fig. 3a-c), consistent with the extrinsic inversion symmetry breaking. The two inversion-breaking routes therefore lead to the same qualitative outcome: centrosymmetric $CsV_3Sb_5$ is diode-inactive when protected, but diode-active when extrinsic symmetry breaking is introduced.

Moreover, we find that the diode response of the inversion-symmetry-broken samples is not a fixed property. In etched Device 13, consecutive current sweeps at the same temperature and nominally identical conditions produce different diode efficiencies. Both the sign and the magnitude of $\eta$ fluctuate from sweep to sweep (Fig. 3a-c, and Fig. S6), although the experimental parameters are unchanged. As the temperature increases from 0.25 to 1.5 K (Fig. 3d&e), the average magnitude decreases, but stochastic polarity and amplitude persist.

The same behavior appears in oxidized samples. Device 12, for example, shows sweep-to-sweep variations of the diode efficiency between −30.86% and 17.07% (Figs. S3 to S5). The effect is reproducible across devices, including a second etched sample (Device 14; Fig. S7), and it is not eliminated by field-cooling followed by removal of the external field (Figs. S8 and S9). These observations are difficult to reconcile with a single fixed structural diode axis or with a static fabrication artifact. Instead, they suggest that the field-free diode signal is a mesoscopic statistical quantity: repeated switching measurements can redistribute the active superconducting paths and reconfigure the internal time-reversal symmetry-broken texture that couples to the extrinsic inversion symmetry breaking.

A decisive feature of the stochastic diode is its progressive stabilization by an out-of-plane magnetic field. In Device 13, zero-field sweeps show random diode polarity, whereas a field of 0.1 T largely fixes the polarity while leaving a broad distribution in amplitude (Fig. 4). At 0.2 T, both the polarity and the amplitude become stable and reproducible. The same field-induced evolution is observed in Device 14 (Figs. S10 and S11). These observations indicate that the external magnetic field selects and stabilizes the preexisting, metastable nonreciprocal state. Control measurements presented in Fig. S12 further rule out ordinary time-dependent fluctuation near the critical current as the origin of the observed stochasticity.

## Control devices without long-range CDW order

The diode response of $CsV_3Sb_5$ requires external inversion symmetry breaking, but inversion symmetry breaking alone is not sufficient. Field-free nonreciprocity also requires a source of time-reversal symmetry breaking. To test whether this source is tied to the CDW-related order in $CsV_3Sb_5$, we performed control measurements on Ti-doped $CsV_{3-x}Ti_xSb_5$ with $x = 0.24$. This doping level suppresses long-range CDW order[55], as confirmed by transport characterization (Figs. S13 and S14). After deliberately introducing inversion-symmetry breaking with the same fabrication protocol used for the pristine devices, the Ti-doped samples do not show a field-free superconducting diode effect.

This control experiment separates extrinsic inversion symmetry breaking from the internal time-reversal symmetry breaking. Oxidation or etching can break the inversion symmetry breaking, but without the CDW-linked order, the field-free diode response does not appear. The result therefore identifies the time-reversal symmetry breaking relevant to the superconducting diode effect as related to the CDW-related electronic orders of the normal states, rather than the standalone superconducting condensate or a trivial by-product of surface processing.

# Discussion

Our measurements define a sequential symmetry test for field-free superconducting nonreciprocity in $AV_3Sb_5$. The first step is the centrosymmetric baseline: hBN-encapsulated, pristine $CsV_3Sb_5$ devices show no detectable field-free superconducting diode effect, consistent with the inversion symmetry of the ideal kagome lattice. The second step is controlled inversion-symmetry breaking: surface oxidation or asymmetric etching activates a sizable field-free diode response. These two observations establish that the polar axis required for nonreciprocal superconductivity is not intrinsic to pristine $CsV_3Sb_5$, but can be supplied extrinsically. Thus, any claim of an intrinsic superconducting diode effect should be approached with caution. The remaining and more revealing observation is that the activated diode is not deterministic. At zero applied field, repeated current sweeps of the same device can yield different diode magnitudes and even opposite polarities. This stochasticity is the key to the physical interpretation.

The sweep-to-sweep stochasticity points to a metastable internal time-reversal symmetry-broken domain configuration rather than a single fixed structural diode axis. A fixed polar axis imposed by oxidation or etching can make a field-free superconducting diode effect symmetry-allowed, but it cannot by itself explain why the diode polarity reverses under nominally identical conditions. Locally, the sign of a superconducting diode contribution is controlled by $(\hat{P} \times \hat{B}_{\text{eff}}) \cdot \hat{I}_c$, where $\hat{P}$ is the local polar axis and $\hat{B}_{\text{eff}}$ denotes the internal time-reversal symmetry-breaking field sensed by the superconducting condensate [3]. In an oxidized or etched thin flake, surface degradation, edges, and disorder naturally break inversion symmetry and induce inhomogeneous superconducting paths. If the internal time-reversal symmetry-broken state is also domain-like, the measured diode efficiency should be understood as a mesoscopic sum,

$$\eta \propto \sum_j w_j \left[ (\hat{P}_j \times \hat{B}_{\text{eff},j}) \cdot \hat{I}_c \right] \tag{2}$$

where $j$ labels superconducting paths or local domains and $w_j$ is the sweep-dependent weight with which each channel contributes to switching. In this picture, consecutive critical-current sweeps do not simply read out a fixed state. By driving parts of an inhomogeneous device through the superconducting-to-resistive transition, each sweep can redistribute current paths and repopulate nearly degenerate time-reversal symmetry-broken domain configurations. A modest rearrangement of either $w_j$ or the signs of $\hat{B}_{\text{eff},j}$ can change the net sign and magnitude of $\eta$.

The magnetic-field training further supports this domain scenario. At zero field, the internal time-reversal symmetry-broken configurations are nearly degenerate, so the net diode response fluctuates from sweep to sweep. A small out-of-plane magnetic field of order 0.1 T largely removes the polarity randomness, indicating that the field biases one orientation of the time-reversal symmetry-broken texture. The amplitude remains broadly distributed at this field, which is natural if the domain polarity has been partially selected while the superconducting current paths and domain weights remain nonuniform. At about 0.2 T, both the sign and magnitude become reproducible, implying that the external field has more completely stabilized the domain configuration that couples to the extrinsic polar axis.

The trainable stochasticity therefore provides transport evidence for time-reversal symmetry-broken domains in the superconducting regime of $CsV_3Sb_5$. The next question is the origin of these domains. $CsV_3Sb_5$ contains no ferromagnetic constituents, and neither oxidation nor etching is expected to generate a permanent macroscopic magnetization. The natural candidate is therefore an orbital, chiral, loop-current-like, or otherwise domain-forming electronic texture associated with the CDW sector, which has been implicated in normal-state time-reversal symmetry breaking by Kerr, muon-spin-rotation, and related measurements[29,31,34]. Upon entering the superconducting state, this CDW-derived time-reversal symmetry-broken texture can be inherited by, or become strongly coupled to, the superconducting condensate. The observed field-free superconducting diode effect then results from coupling this internal time-reversal symmetry-broken domain landscape to the extrinsic inversion breaking introduced by oxidation or etching.

The Ti-doped control experiment corroborates the CDW scenario by removing the CDW ingredient while keeping the symmetry-breaking fabrication route. In $CsV_{3-x}Ti_xSb_5$ with $x$ = 0.24, long range CDW order is suppressed[55]. After applying the same oxidation or etching protocols that break inversion symmetry in pristine $CsV_3Sb_5$, the Ti-doped devices still show no field-free superconducting diode effect. Thus, extrinsic polarity, fabrication disorder, and inhomogeneous switching are not sufficient. The missing ingredient is the CDW-linked time-reversal symmetry broken electronic texture. This control experiment indicates that the time-reversal symmetry breaking detected through superconducting-state nonreciprocity is rooted in the CDW-related normal-state order and is carried into the superconducting phase through their coupling.

Taken together, the experiments define a symmetry-and-control logic for field-free nonreciprocity in $CsV_3Sb_5$. The absence of an intrinsic superconducting diode effect in pristine, centrosymmetric devices establishes that inversion symmetry blocks the diode response even if internal time-reversal symmetry breaking is present. The emergence of a finite response after oxidation or asymmetric etching shows that the time-reversal symmetry is still broken. The sweep-to-sweep stochasticity and its stabilization by out-of-plane magnetic field then point to metastable time reversal symmetry-broken domain configurations. Finally, the disappearance of the response in similarly processed Ti-doped $CsV_{3-x}Ti_xSb_5$ suggests that the TRS breaking in the superconducting states is associated with CDW-related orders. Our findings not only uncover a stochastic, field-free superconducting diode effect in engineered kagome superconductors and reveal its time-reversal symmetry breaking origin, but also establish a framework for engineering controllable, low-dissipation nonreciprocal superconducting devices.

# Material and Method

## Device fabrication

High-quality $CsV_3Sb_5$ flakes were mechanically exfoliated in an argon-filled glovebox with $O_2$ and $H_2O$ levels below 0.01 ppm. Pristine devices were fabricated using two different approaches. For bottom-electrode devices, Pd/Au electrodes were first patterned on the substrate, followed by dry transfer of $CsV_3Sb_5$ flakes onto the pre-patterned electrodes and subsequent encapsulation with hBN. For top-electrode devices, $CsV_3Sb_5$ flakes were first exfoliated onto the substrate, followed by electron-beam lithography (EBL), Pd/Au deposition by magnetron sputtering, and lift-off to define the electrodes, after which the devices were encapsulated with hBN. For oxidation experiments, the hBN layer was removed and devices were exposed to ambient air.

## Transport measurements and diode-efficiency analysis

Electrical transport measurements were performed in both a Quantum Design Physical Property Measurement System (PPMS) and an Oxford Instruments Triton XL1000 dilution refrigerator. The Differential resistance ($dV/dI$) was measured using a combined DC and AC current bias. A DC bias current was supplied by a Yokogawa GS200 current source, while a small AC excitation current was applied and the corresponding voltage response was detected using a Stanford Research Systems SR830 lock-in amplifier.

## Data availability

The data that support the plots within this paper and other related findings are available from the corresponding author upon reasonable request.

## Acknowledgments

Ben-Chuan Lin thanks Manfred Sigrist for valuable discussions. This work was supported by the National Key Research and Development Program of China (Grants No. 2022YFA1204004 and No. 2022YFA1403700), the National Natural Science Foundation of China (Grants No. 52071026, 52130103, and 12374099), the Guangdong Basic and Applied Basic Research Foundation (Grants No. 2022B1515130005 and 2023A1515140171), Quantum Science and Technology–National Science and Technology Major Project (Grants No. 2021ZD0303000 and 2021ZD0303001), and Shenzhen International Quantum Academy Grants (No. SIQA2024KFKT03).

## Author contributions

B.-C.L. supervised the project, conceived the study, and designed the experiments.

X.-J.L., S.W., K.-F.S., S.S., L.X. and B.-C.L. fabricated the microdevices and performed the transport measurements, with the assistance of Y.-Q.B., N.K., X.W., Y.F.W., S.G.W. and D.P.Y.

T.L. and Q.L.H. synthesized the doped single crystals.

J.-P.P., X.-J.L. and K.-F.S. performed the AFM measurements.

X.F. and K.T.L. contributed to the theoretical analysis.

B.-C.L., X.-J.L., S.W. and X.F. wrote the manuscript with the necessary input from all authors.

## Competing interests

All authors declare they have no competing interests.

# Figures

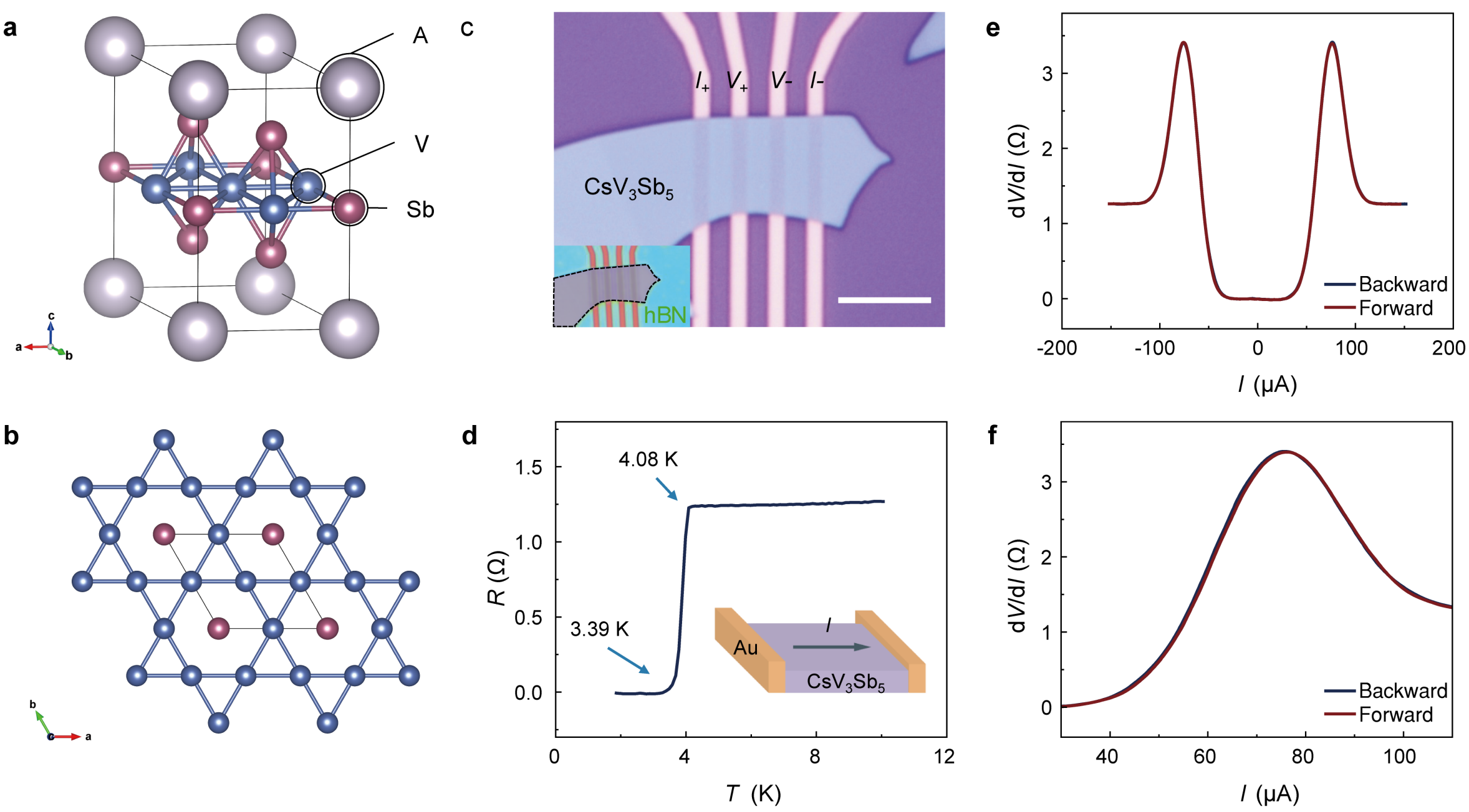

**Fig. 1. Pristine encapsulated $CsV_3Sb_5$ devices exhibit no field-free superconducting diode effect. a** Crystal structure of $AV_3Sb_5$. The kagome lattice is formed by V atoms (blue), while Sb atoms (pink) occupy two distinct sites. **b** The V–Sb sublattice forms a kagome network, with V atoms arranged in a corner-sharing triangular lattice and Sb atoms located at the centers of the kagome hexagons. **c** Optical micrograph of $CsV_3Sb_5$ Device 1. The inset shows the same device encapsulated with hBN. Current is applied through $I_+$ and $I_-$, and voltage is measured between $V_+$ and $V_-$. Scale bar, 5 $\mu$m. **d** Resistance versus temperature for Device 1. Left arrows mark $T_c^{onset}$ = 4.08 K and $T_c^{zero}$ = 3.39 K. Lower-right inset: schematic of the measurement configuration. **e&f** Differential resistance $dV/dI$ as a function of DC bias current $I$ at $T$ = 1.8 K. In **f**, the negative-bias branch of the backward sweep is inverted for direct comparison with the forward sweep, demonstrating the absence of asymmetry between $|I_c^+|$ and $|I_c^-|$.

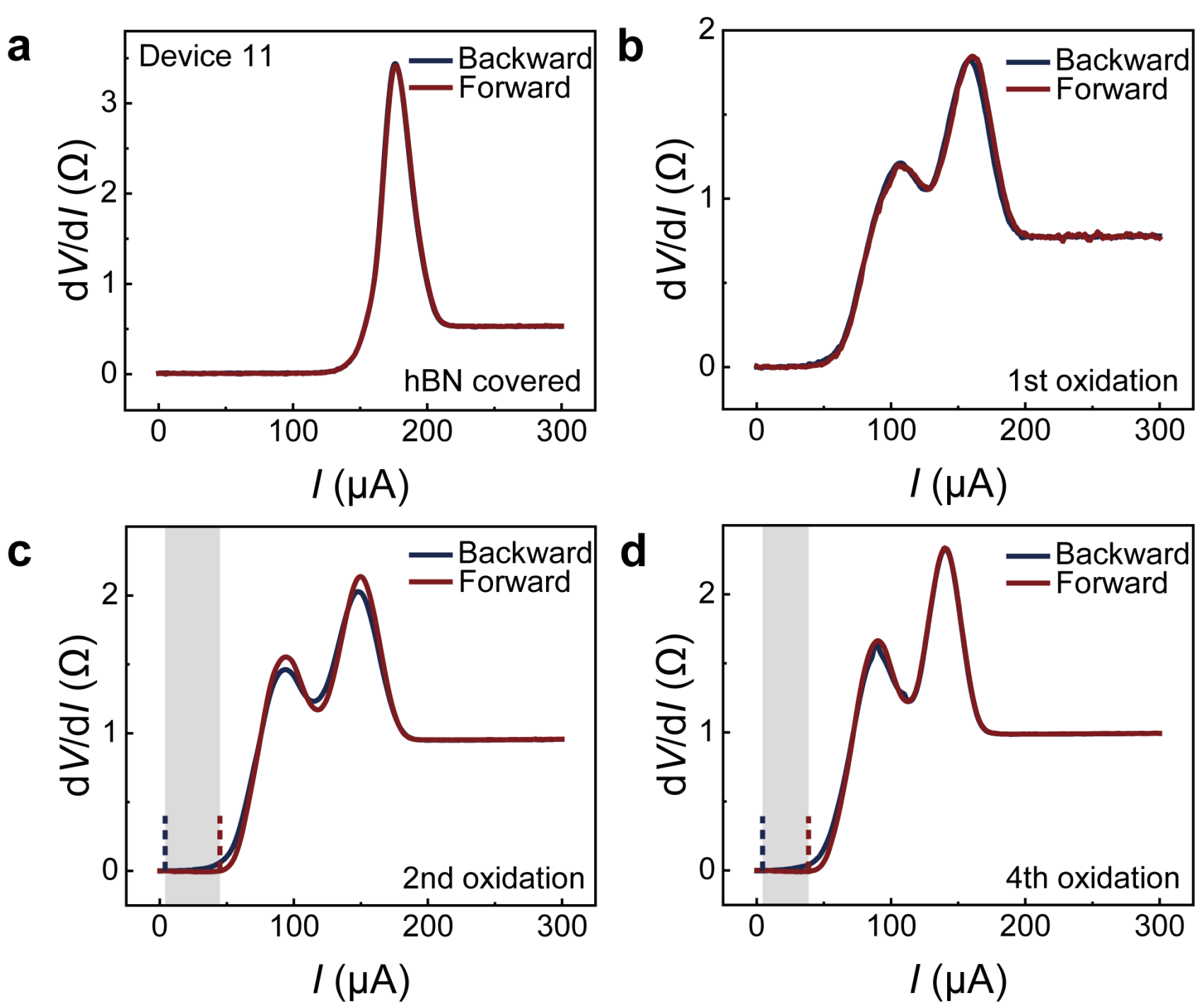

**Fig. 2. Surface oxidation induces a field-free diode response by breaking inversion symmetry.** Evolution of the differential resistance $dV/dI$ as a function of DC bias current $I$ in oxidized $CsV_3Sb_5$ Device 11 at $T = 1.8$ K and $B = 0$ T. The forward sweep is from 0 to +300 $\mu$A and the backward sweep is from 0 to -300 $\mu$A; the backward branch is inverted for comparison. **a** Pristine hBN-covered device. **b** Differential resistance after removal of the hBN and surface oxidation. **c** Device after the second oxidation step. Red and blue short-dashed lines indicate $|I_c^+|$ and $|I_c^-|$, and the shaded region marks $||I_c^+| - |I_c^-||$. **d** Device after the fourth oxidation.

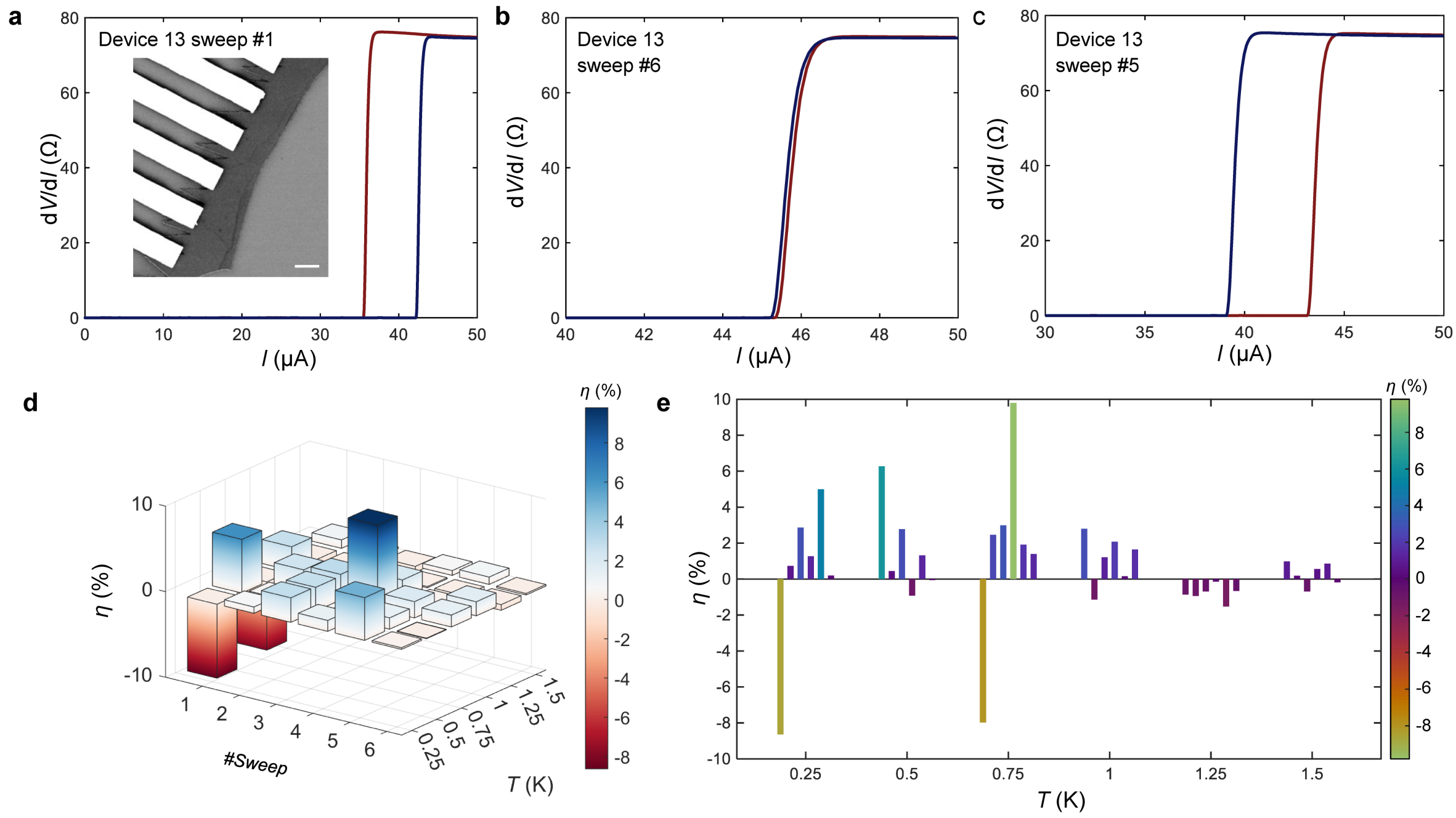

**Fig. 3. Geometrically inversion-broken devices show stochastic field-free superconducting diode behavior. a-c** Differential resistance $dV/dI$ as a function of DC bias current $I$ in the etched zig-zag Device 13. The inset in **a** shows a scanning electron microscopy image of the etched device; Scale bar, 1 $\mu$m. Red and blue curves denote the forward and backward sweeps, respectively, with the backward branch inverted for comparison. **d&e** Statistical distribution of the diode efficiency as a function of temperature. Measurements were performed from 0.25 to 1.5 K in 0.25 K steps, with six consecutive sweeps taken at each temperature. The average diode amplitude decreases with increasing temperature, but stochastic variation persists.

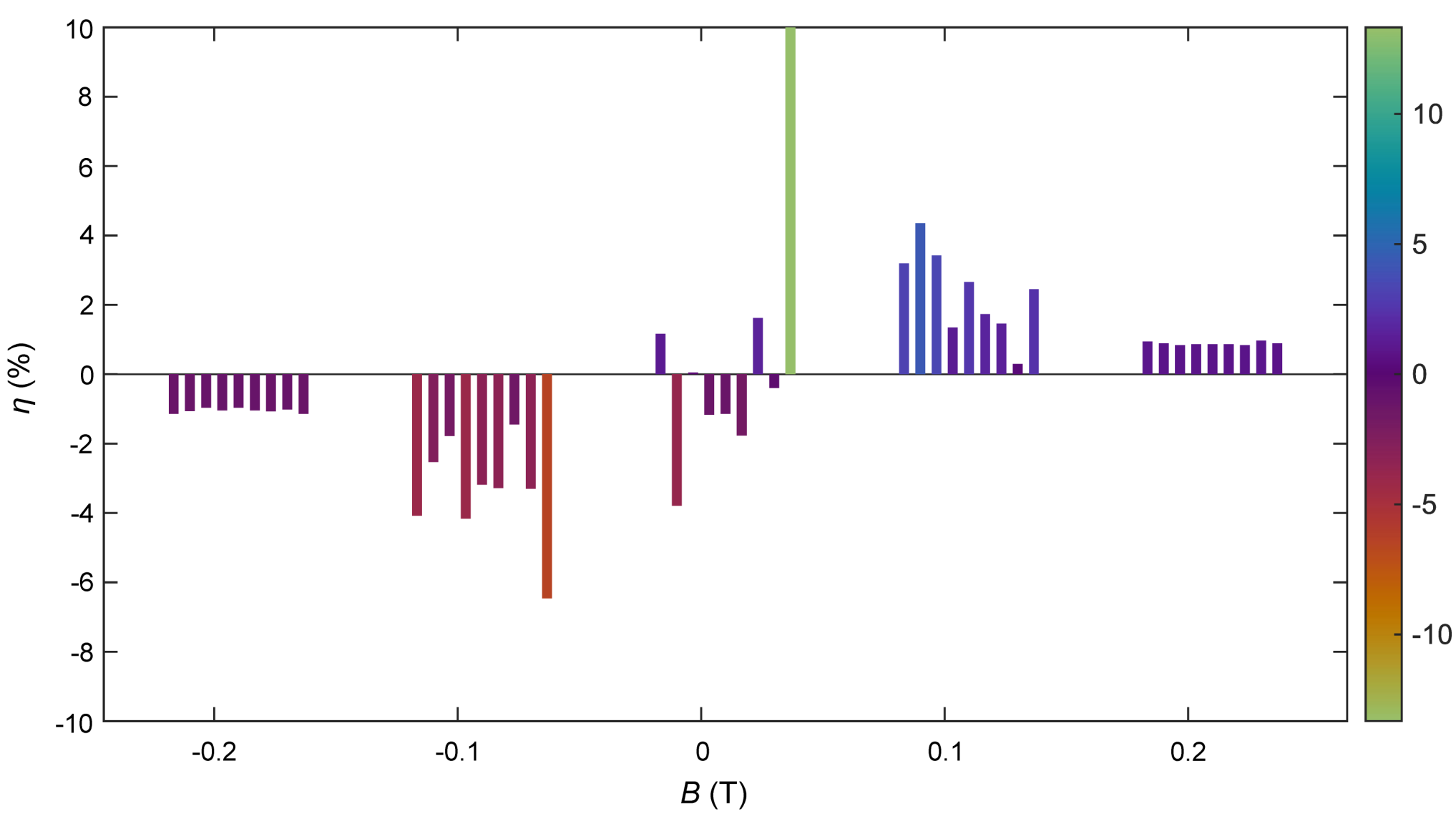

**Fig. 4. Out-of-plane magnetic fields stabilize the stochastic superconducting diode response.** Diode efficiency of Device 13 under magnetic fields between -0.2 and 0.2 T, with nine consecutive sweeps at each field. The field fixes the diode polarity and, at larger magnitude, stabilizes the amplitude.

# Supplementary Materials for

## Symmetry Origins of the Field-Free Superconducting Diode Effect in the Kagome Superconductor $CsV_3Sb_5$

Xin-Jie Liu *et al.*

*Corresponding author. Email: yanfeiwu@ustb.edu.cn; sgwang@ahu.edu.cn; wangshuo@iqasz.cn; linbenchuan@iqasz.cn

**This PDF file includes:**

**Supplementary Text**

In this study, $CsV_3Sb_5$ samples were fabricated with either top electrodes (Fig. S1) or bottom electrodes (Fig. S2), and all measurements were performed with hBN encapsulation. Device thicknesses range from 21 to 76 nm, with critical currents from 81 μA to 496.5 μA. Overall, the statistics suggest that all devices with different fabrication methods show no discernible superconducting diode effect (SDE).

For each oxidation stage, the differential resistance $dV/dI$ as a function of applied DC bias current $I$ was measured repeatedly at 1.8 K. During the second oxidation stage, three consecutive measurements were performed, and only the second exhibited a field-free SDE, corresponding to Fig. 2(c) in the main text. In the third oxidation stage, three additional measurements were carried out, none of which showed SDE. During the fourth oxidation stage, three additional sweeps were taken, with the second again displaying a field-free SDE, corresponding to Fig. 2(d) in the main text. At the final oxidation stage, all three sweeps showed a monotonic increase of the resistance in the backward branch within the 0~50 μA range. No discernible critical current was observed. This indicates that superconductivity is completely suppressed. These results demonstrate that the field-free SDE during oxidation is unstable and fluctuates between repeated sweeps, which will be further investigated in Figs. 3&4.

The differential resistance $dV/dI$ as a function of applied DC bias current $I$ was measured repeatedly on another Device 13 at 1.8 K over six successive sweeps, as shown in Fig. S4. Field-free SDE is observed in sweeps #1, #5, and #6 (panels a, e, and f), whereas sweeps #2, #3, and #4 (panels b, c, and d) do not exhibit SDE. The extracted SDE efficiency $\eta$ is 17.07% for sweep #1, -30.86% for sweep #5, and -14.07% for sweep #6, reflecting the stochastic nature of the oxidation-induced field-free SDE.

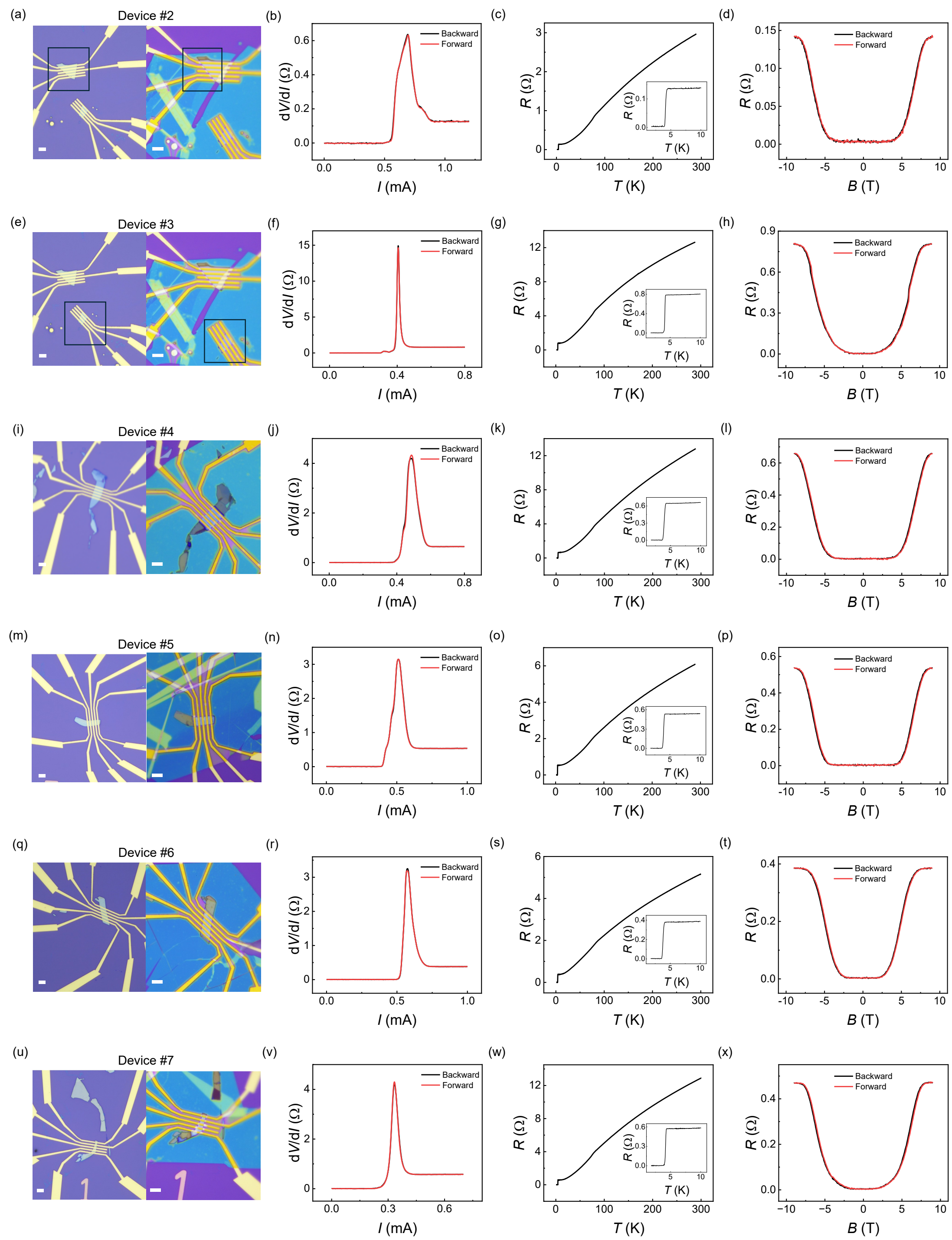

**Fig. S1. The pristine $CsV_3Sb_5$ devices with electrodes evaporated on top surfaces.** All measurements were performed with the sample encapsulated by hBN. A 5-µm scale bar is labeled in the lower-left corner of the figure. Also shown are the differential resistance as a function of bias current at 1.8 K, the zero-field resistance–temperature curve, and the resistance as a function of in-plane magnetic field at 1.8 K. The extracted critical current $I_c$, critical temperature $T_c$, critical magnetic field $B_{c\parallel}$, and the upper bound of the SDE efficiency $\eta$ are summarized in Table 1.

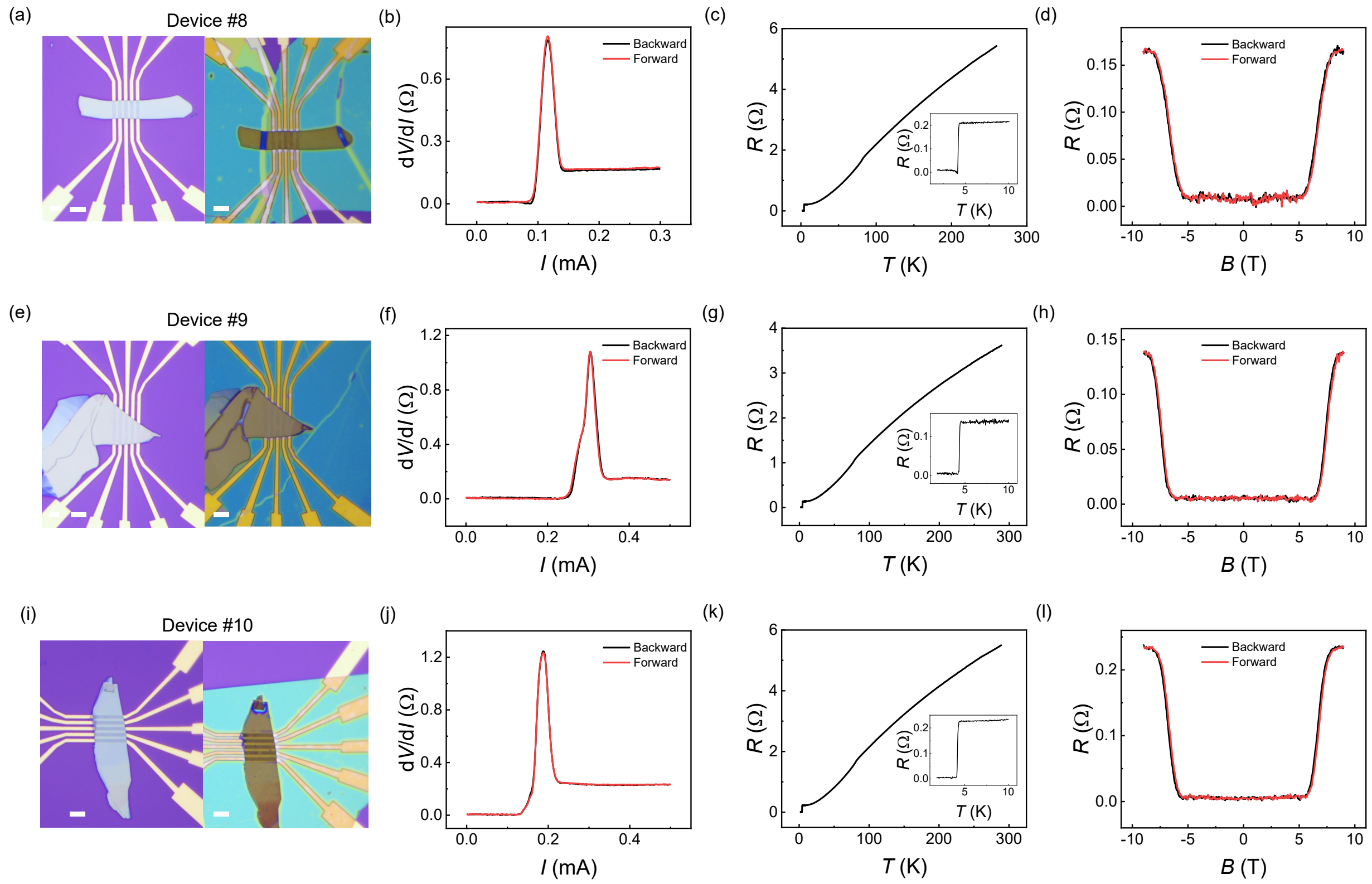

**Fig. S2. The pristine $CsV_3Sb_5$ devices, where $CsV_3Sb_5$ has been directly transferred onto pre-patterned electrodes.** All measurements were performed with the device encapsulated by hBN. The scale bar is 5 μm. Also shown are the differential resistance d$V$/d$I$, the resistance–temperature curve, and the magnetoresistance under in-plane magnetic field at 1.8 K. The extracted critical current $I_c$, critical temperature $T_c$, critical magnetic field $B_{c\parallel}$, and the upper bound of the SDE efficiency $\eta$ are summarized in Table 1.

**Table S1.**

Extracted superconducting parameters from $CsV_3Sb_5$ devices shown in Figure S1&2.

| Device | $d$ (nm) | $I_c$ (μA) | $T_c$ (K) | $B_{c\parallel}$ (T) | $\eta$(%) |
|---|---|---|---|---|---|
| **2** | 65.15 | 490 | 4.12 | 6.51 | 0.2163 |
| **3** | 23.34 | 300 | 4.12 | 5.99 | 0 |
| **4** | 21.06 | 370.5 | 3.96 | 6.69 | 0.1028 |
| **5** | 59.54 | 381 | 3.95 | 6.51 | 0 |
| **6** | 44.90 | 496.5 | 3.80 | 4.97 | 0.4806 |
| **7** | 62.65 | 214.4 | 4.17 | 5.89 | 0 |
| **8** | 63.25 | 81.4 | 4.20 | 6.69 | 0 |
| **9** | 75.56 | 228.4 | 4.36 | 7.49 | 0.0328 |
| **10** | 46.72 | 129.4 | 4.17 | 6.77 | 0.0532 |

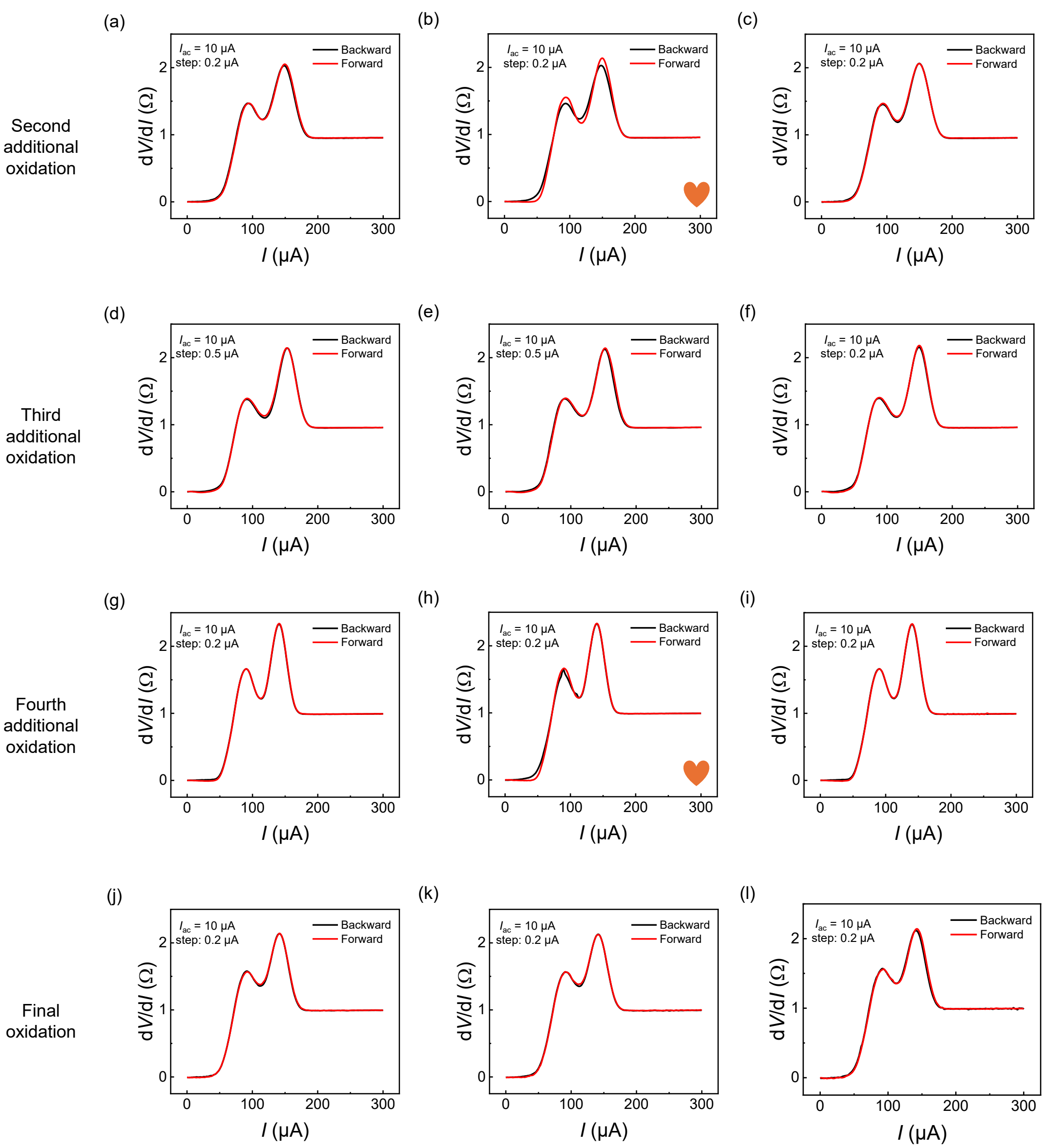

**Fig. S3. Differential resistance of Device 12 (Fig. 2 in the main text) after successive oxidation stages.** Data acquired at 1.8 K are shown as a function of applied DC bias current for the (a–c) second, (d–f) third, (g–i) fourth, and (j–l) final oxidation stages. For each oxidation cycle, the system was heated to 300 K and vented, and the sample was subsequently taken out and exposed to ambient air to promote surface oxidation.

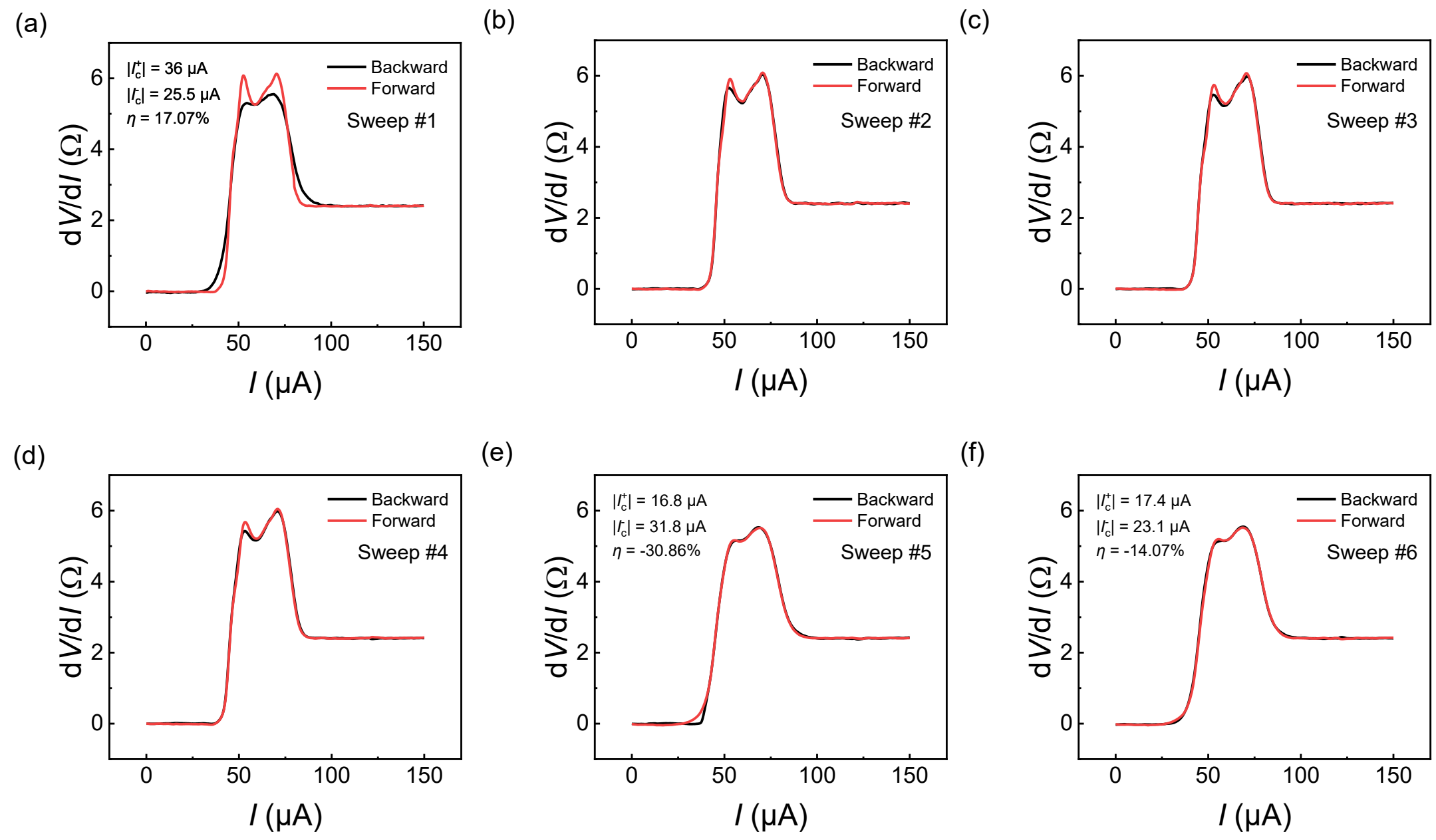

**Fig. S4. Differential resistance of Device 13 after successive oxidation stages.** Differential resistance measured at 1.8 K as a function of applied DC bias current. Panels (a-f) correspond to Sweep #1 through Sweep #6, respectively. Panels (a), (e), and (f) present field-free SDE; the extracted $I_c^+$, $I_c^-$, and SDE efficiency $\eta$ are indicated in the upper left corner of each panel.

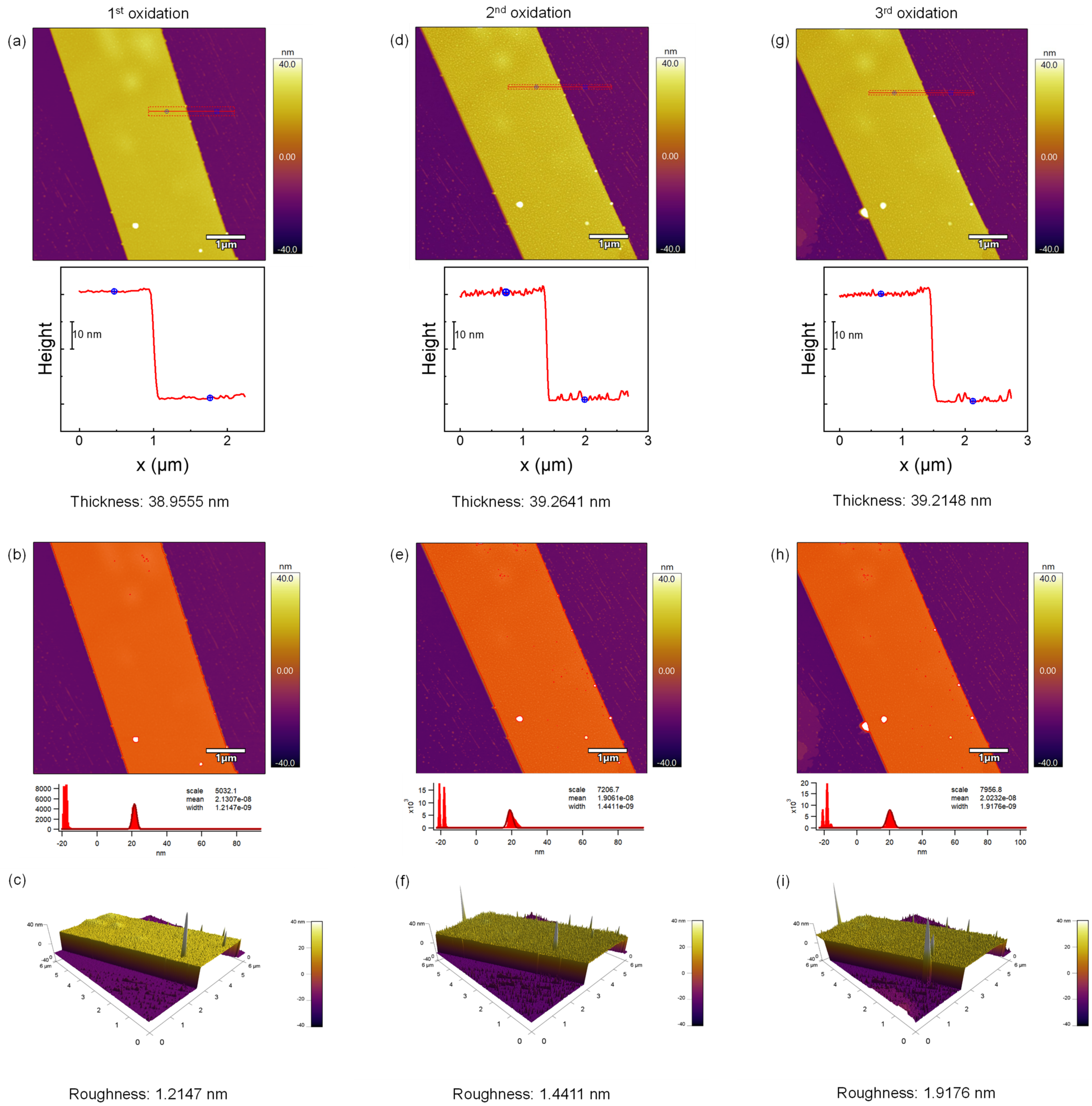

**Fig. S5. Atomic force microscopy (AFM) characterization before and after successive oxidation.** AFM characterization of the surface of our sample, and following the first, second and third additional oxidation steps, including height line profiles (a, d, g), roughness (b, e, h) and 3D AFM images (c, f, i).

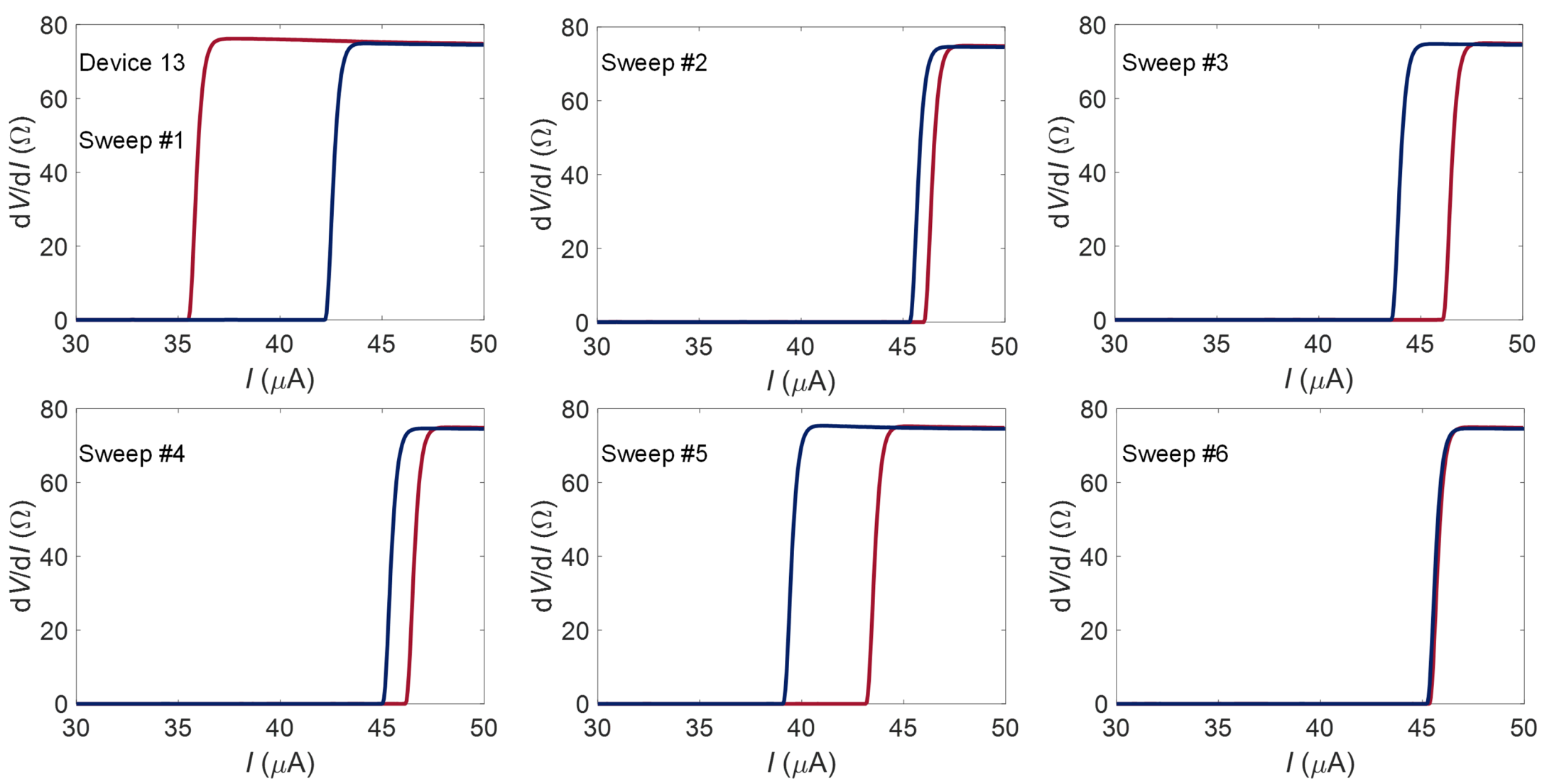

**Fig. S6. Different current sweeps of the SDE in Device 13.**

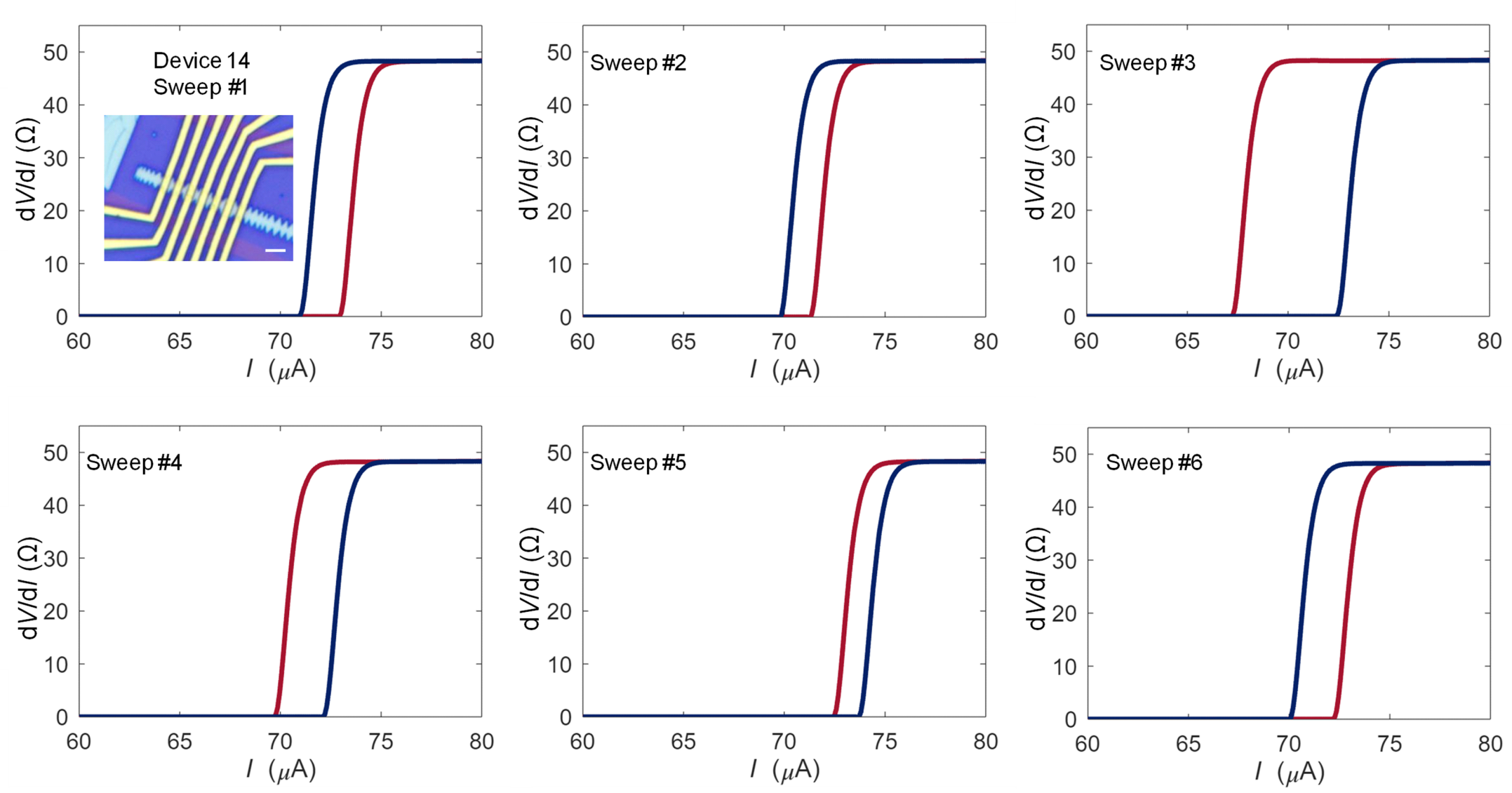

**Fig. S7. The SDE of a new etched Device 14 with a well-defined superconducting branch.** Here the scale bar is 3 μm. The sweep rate of the bias current is 0.05 μA/s.

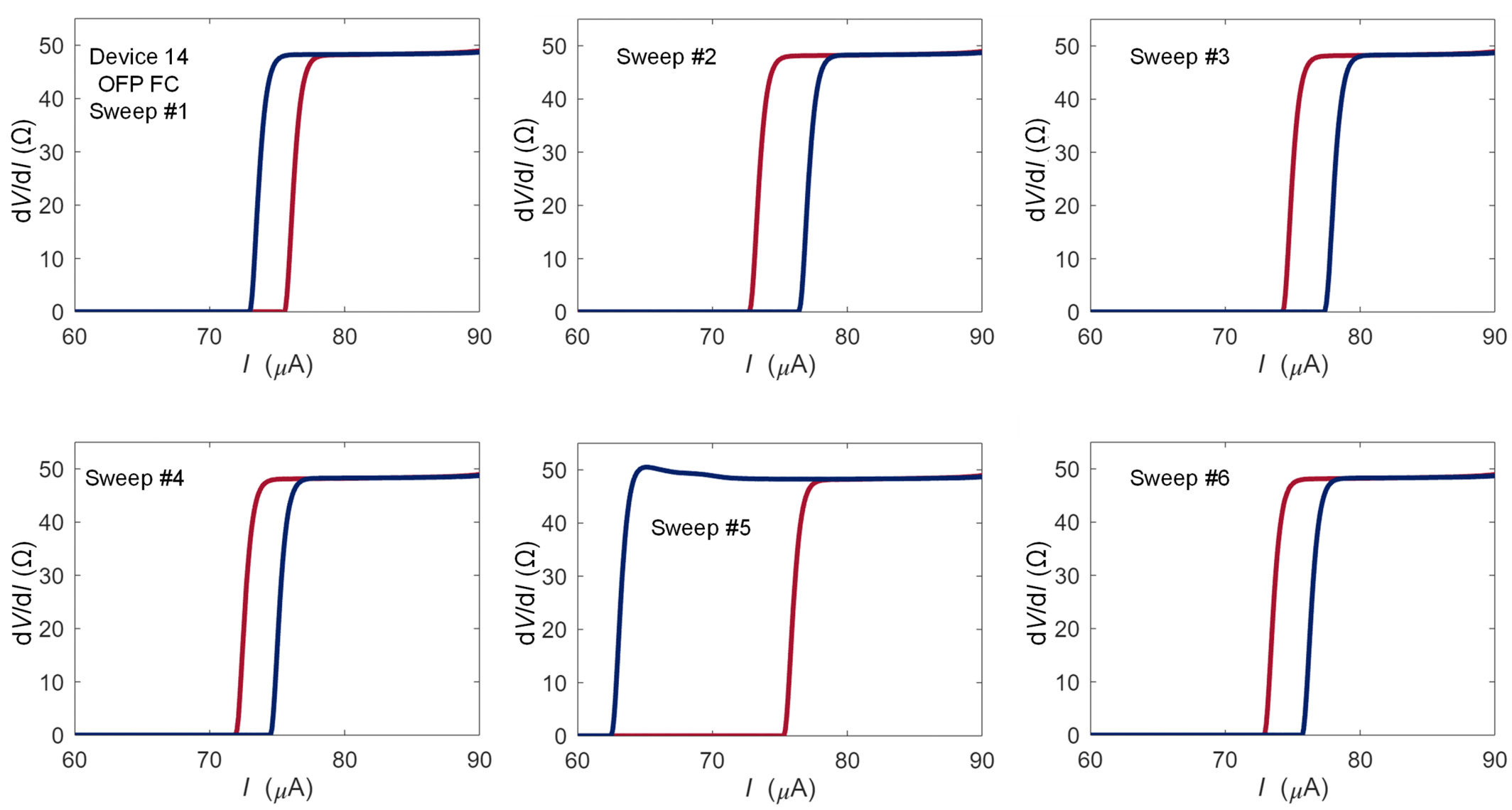

**Fig. S8. Out-of-plane field-cooling measurement of Device 14.** The field-cooling procedure was performed by cooling the sample to the base temperature under an external out-of-plane magnetic field of 1 T, followed by removal of the field before the SDE measurement. The stochastic SDE still persists under this protocol.

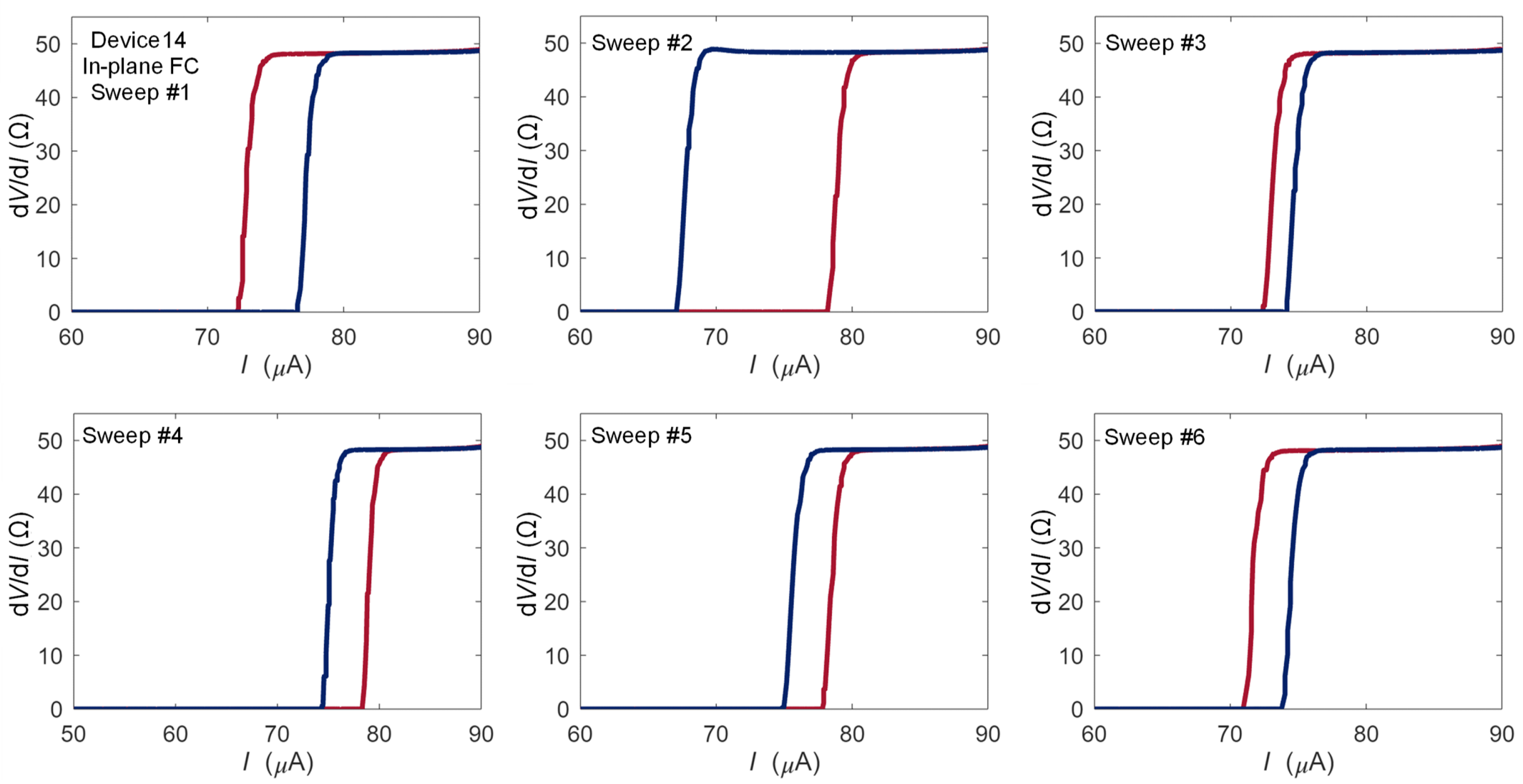

**Fig. S9. In-plane field-cooling measurement of Device 14.** The field-cooling procedure was performed by cooling the sample to the base temperature under an external in-plane magnetic field of 1 T, followed by removal of the field before the SDE measurement. It can be seen that the stochastic SDE still persists under this protocol.

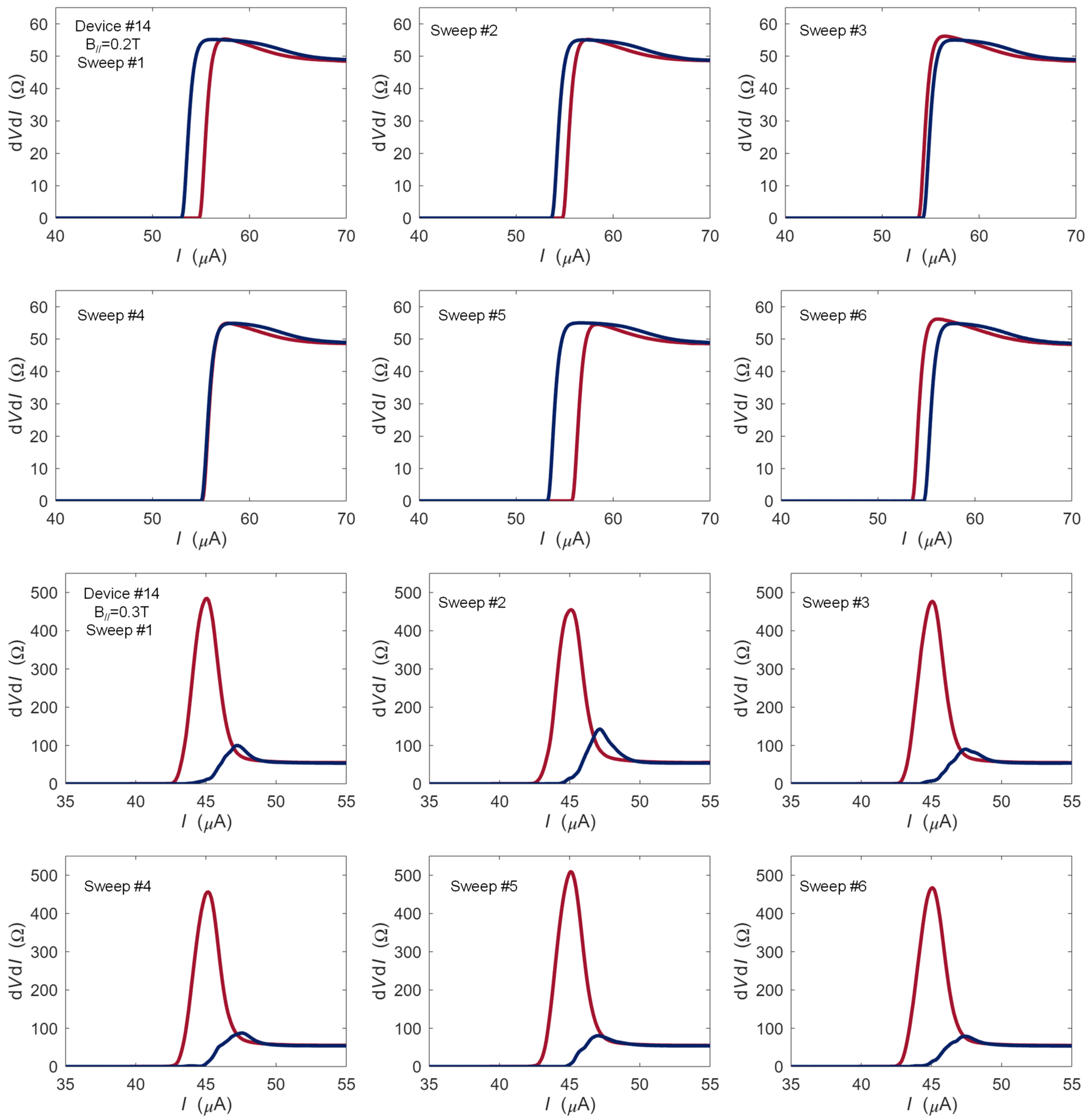

**Fig. S10. An in-plane magnetic field of 0.3 T stabilize the SDE of Device 14.** The sweep rate is 0.05 μA/s. For an in-plane magnetic field of 0.2 T, the SDE is still stochastic. The in-plane field of 0.3 T is below the characteristic field scale for vortex entry in this geometry. Using the measured out-of-plane critical field $\mu_0 H_{c2}^{\perp}$ of ~ 0.85 T, the coherence length is estimated to be approximately 19.6 nm, according to the formula $\xi = \sqrt{\Phi_0/(2\pi\mu_0 H_{c2}^{\perp})}$. For a sample thickness $d$ of approximately 40 nm, the corresponding parallel lower critical field $H_{c1}$ (rather than $H_{c2}$) is estimated to be 0.59 T, according to the formula $\mu_0 H_{c1}^{\parallel} = \frac{2\Phi_0}{\pi d^2} \ln(\frac{d}{\xi})$, which is much larger than the in-plane training field of 0.3 T. This means that ordinary vortex entry into the superconducting layer is not expected under the in-plane training condition, demonstrating that the field stabilization is not related to the vortex physics.

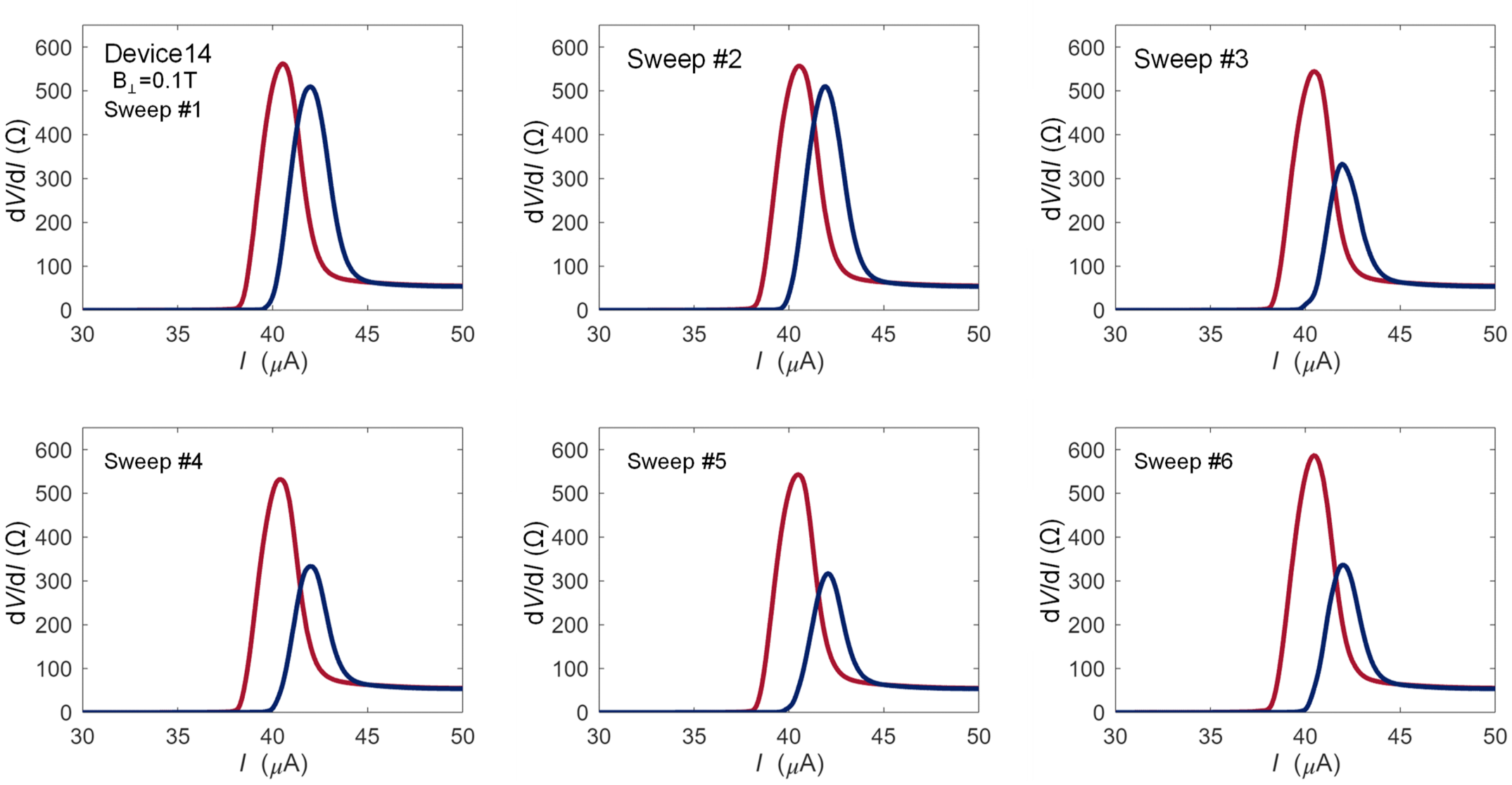

**Fig. S11. An out-of-plane magnetic field of 0.1 T would stabilizes the SDE of Device 14.** The sweep rate is 0.05 μA/s.

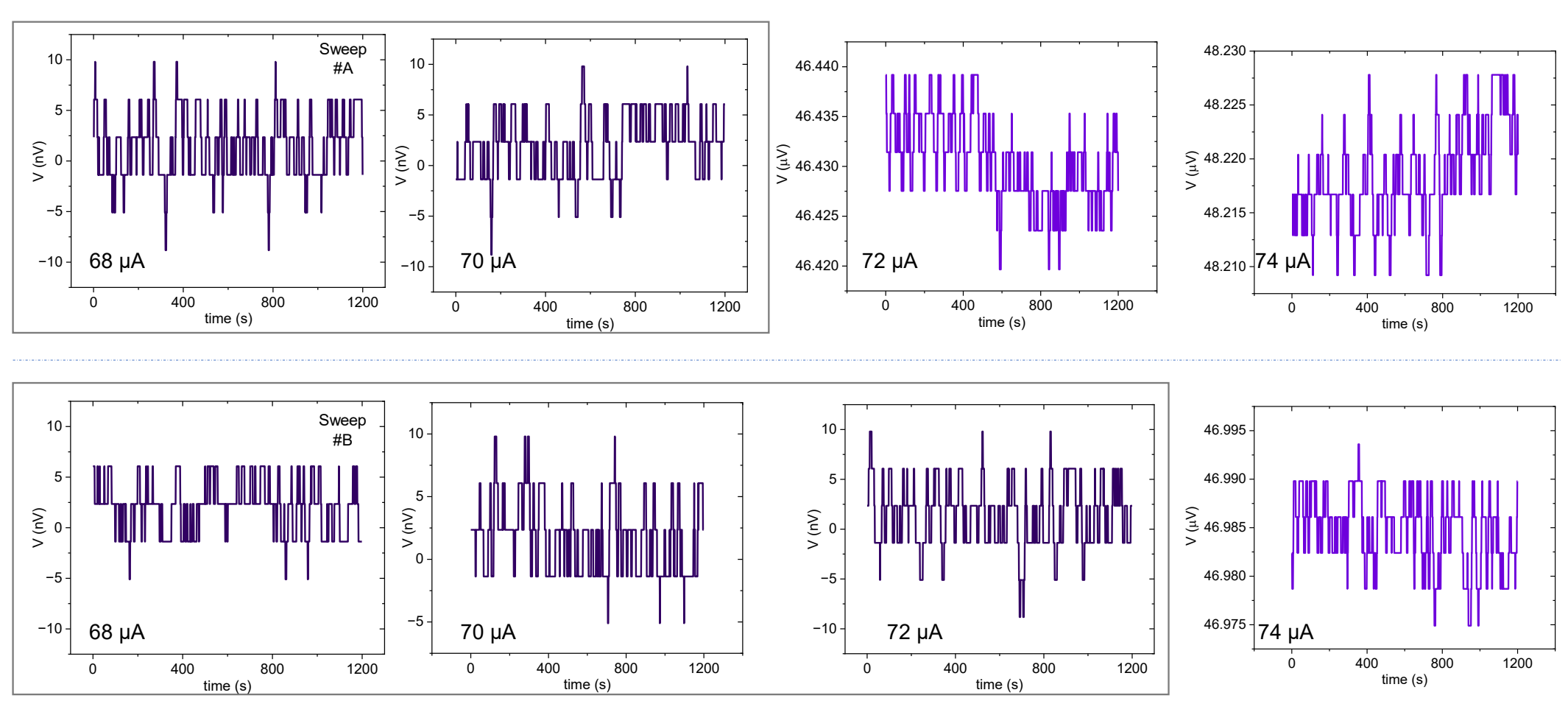

**Fig. S12. Time-dependent fluctuation measurements of two representative sweeps of Device 14.** Owing to the stochastic nature of the superconducting diode effect, the transition region varies from sweep to sweep. The stochastic diode behavior in the main text appears over a broad critical-current window ranging from approximately 66 to 74 μA, corresponding to a total span of about 8 μA. To probe time-dependent fluctuations near the superconducting-to-normal transition, we use a 2 μA current interval, which is much smaller than the total span of 8 μA. For different sweeps (#A and #B), the critical current may shift, and thus the superconducting state corresponds to different bias currents in each sweep. The black curves denote the superconducting state, while the purple curves represent the normal states. Within this fixed-bias measurement window, we do not observe evidence of fluctuations that could account for the full stochastic variation of the diode polarity. In addition, the effective current-sweep time associated with the 8 μA stochastic window is about 3 minutes at the sweep rate used in the main text. To better examine the time-dependent fluctuation effect, the duration of the time-resolved measurement is about 20 minutes, which is substantially longer than the stochastic window. If the stochastic polarity change were primarily caused by simple time-dependent fluctuations or telegraph-noise-like switching at a fixed bias, one would expect this to manifest clearly within the time traces near the transition. Thus, our time-dependent fluctuation measurements rule out such a scenario.

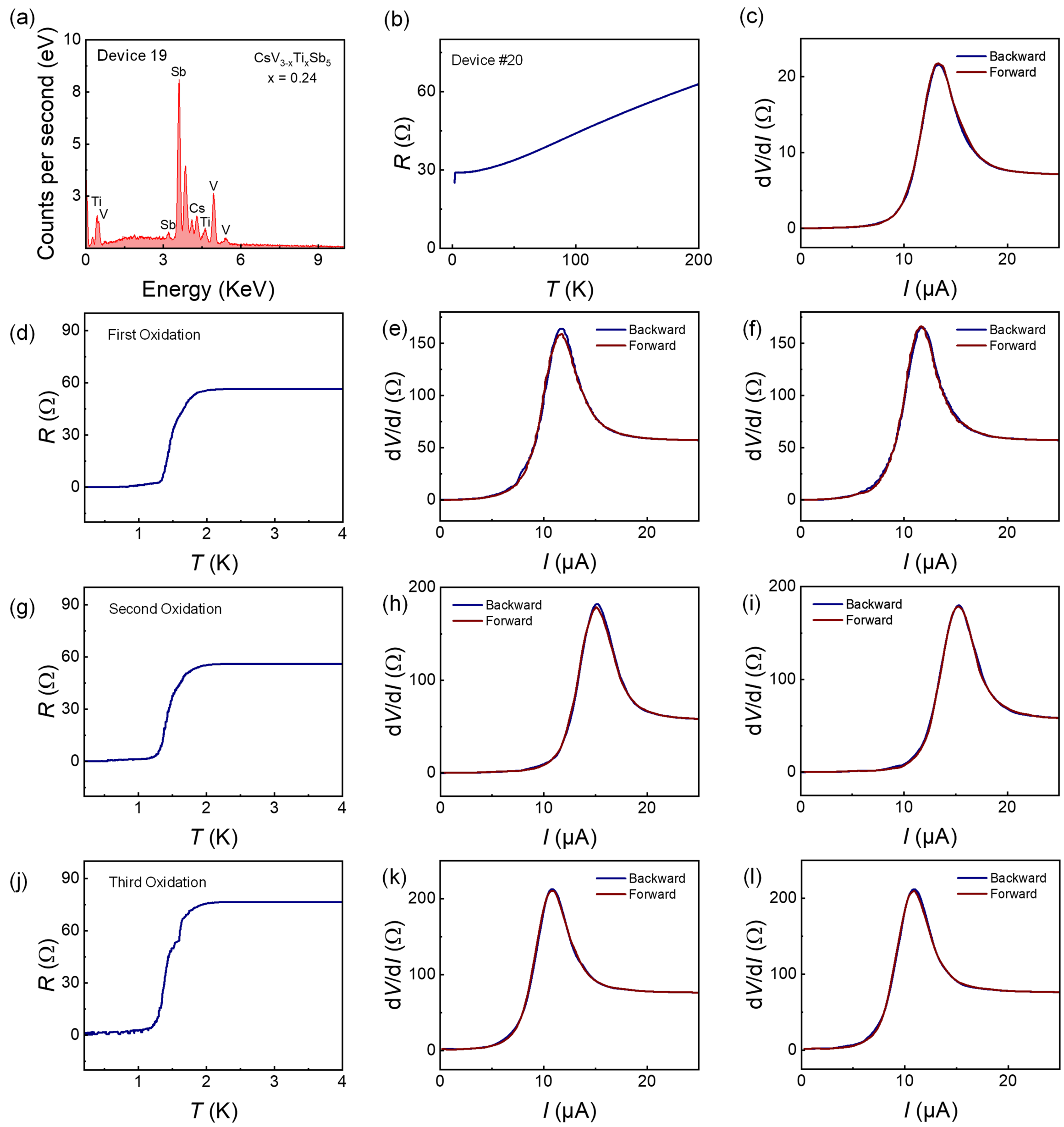

**Fig. S13. Absence of SDE in the Ti-doped sample $CsV_{3-x}Ti_xSb_5$, with $x = 0.24$ (Device 15).** (a) Energy-dispersive X-ray spectroscopy (EDS) analysis of the doped sample. (b) Resistance-Temperature (*R*-*T*) curve in the normal-state temperature range. No signature of a CDW transition is observed, consistent with previous literature (PHYSICAL REVIEW B 112, 144512 (2025)), which shows that such a large Ti concentration ($x > 0.12$) suppresses the CDW order. (c) No SDE is observed, with the forward and backward current sweeps fully overlapping. (d-f) Results after the first oxidation process, following the same procedure described in the main text, still show no observable SDE. (g-i) Results after the second oxidation process. (j-l) Results after the third oxidation process.

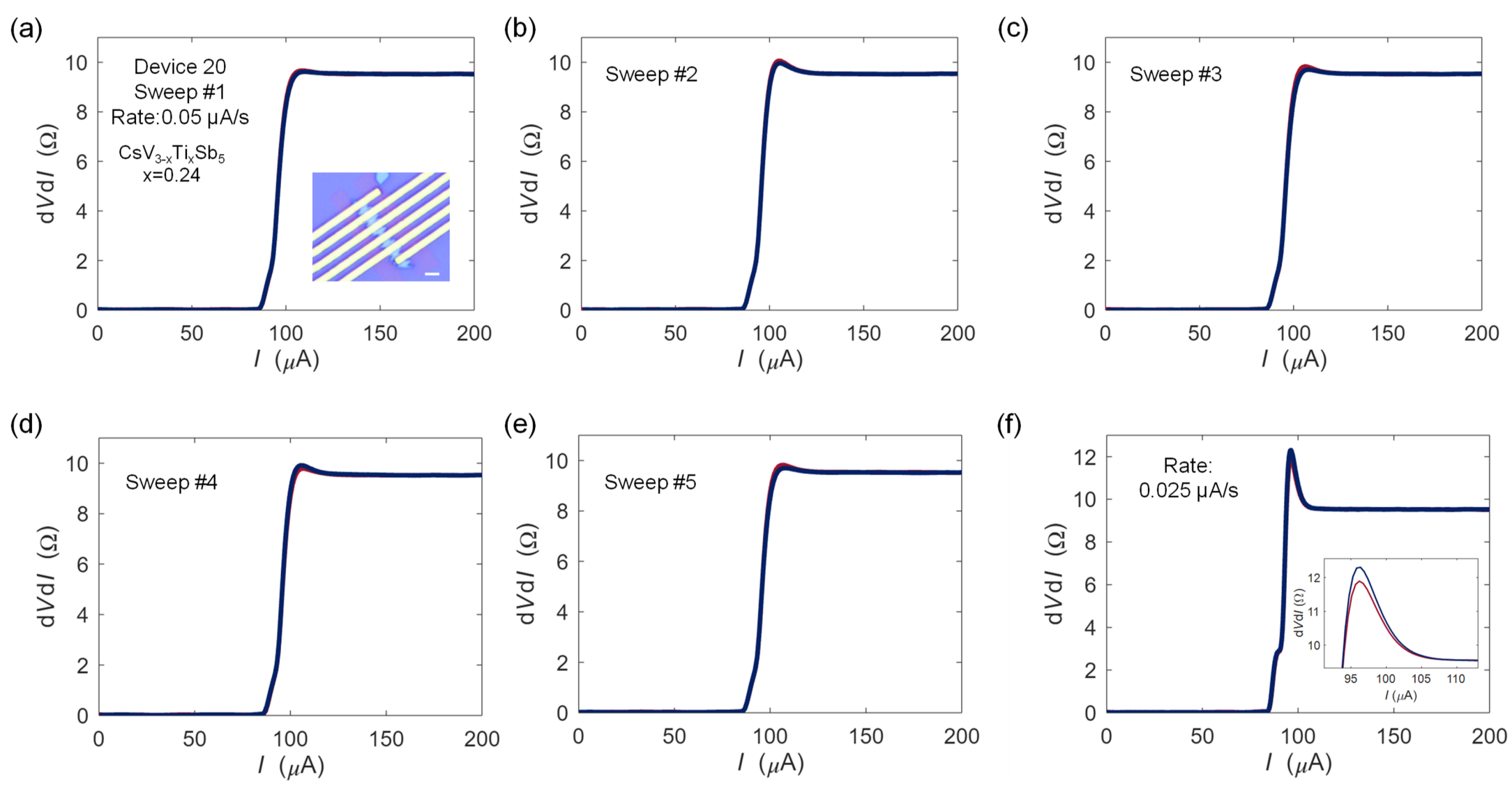

**Fig. S14. Absence of SDE in the etched Ti-doped sample $CsV_{3-x}Ti_xSb_5$, with $x$ = 0.24 (Device 16).** (a-e) No SDE is observed in the etched Ti-doped sample following the same procedure described in the main text, with the forward and backward current sweeps fully overlapping. Inset: Optical image of the etched sample. The scale bar is 3 µm. (f) With a much slower sweep rate of the bias current, no SDE is clearly observed. Overall, the upper bound of the SDE efficiency is approximately 0.1%.